\documentclass[reprint,amsmath,amssymb,aps]{revtex4-1}

\usepackage{graphicx}
\usepackage{dcolumn}
\usepackage{bm}
\usepackage{hyperref}
\usepackage{epstopdf}
\usepackage{float}
\makeatletter\renewcommand{\maketag@@@}[1]{\hbox{\m@th\normalsize\normalfont#1}}
\usepackage{color, soul}

\begin{document}

\preprint{APS/123-QED}

\title{Spiral Transformation for High-Resolution and Efficient Sorting of Optical Vortex Modes}

\author{Yuanhui Wen$^{1,\dagger}$, Ioannis Chremmos$^{1,2,\dagger}$, Yujie Chen$^{1,*}$, Jiangbo Zhu$^3$, Yanfeng Zhang$^1$, and Siyuan Yu$^{1,3,*}$}

\affiliation{$^1$State Key Laboratory of Optoelectronic Materials and Technologies, School of Electronics and Information Technology, Sun Yat-sen University, Guangzhou 510275, China
\\$^2$Hellenic Electricity Distribution Network Operator S. A., Athens, Greece
\\$^3$Photonics Group, Merchant Venturers School of Engineering, University of Bristol, Bristol BS8 1UB, UK
\\$^{\dagger}$These authors contributed equally to this work
\\$^*$Corresponding authors: chenyj69@mail.sysu.edu.cn; s.yu@bristol.ac.uk}

\begin{abstract}
Mode sorting is an essential function for optical systems exploiting the orthogonality of photonic orbital angular momentum mode space. The familiar log-polar optical transformation provides an efficient yet simple approach, however with its resolution restricted by a considerable overlap between adjacent modes resulting from the limited excursion of the phase along one complete circle around the optical vortex axis that is transformed into a line segment with finite length. We propose and experimentally verify a new optical transformation that maps spirals instead of closed concentric circles to parallel lines. As the phase excursion along a spiral in the transverse plane of orbital angular momentum modes is theoretically unlimited, this new optical transformation separates vortex modes with superior resolution while maintaining unity efficiency.
\end{abstract}

\maketitle

Light carrying orbital angular momentum (OAM), also known as optical vortices, possesses a helical phase structure in the form of $\exp(i\ell \theta )$, where $\theta$ is the azimuthal angle around the beam axis and integer $\ell$ is the topological charge which translates to OAM of $\ell \hbar$ per photon \cite{PhysRevA.45.8185}. Either in the form of free space vortex beams or optical fiber modes, OAM states of light with different topological charges are mutually orthogonal and thus constitute a high-dimensional state space that can be exploited to boost the information capacity in both classical \cite{Gibson:04,Nat.Photon.6.488,Bozinovic1545,Nat.Common.5.4876} and quantum communication systems \cite{PhysRevA.78.062320,PhysRevA.64.012306,Nature412.313,Leach662}.

In addition to generating these OAM modes \cite{BEIJERSBERGEN1994321,Heckenberg:92,Cai363}, a critical component of every OAM-based photonic information system is an OAM mode sorter capable of efficiently (i.e., without losing optical energy) separating different OAM modes with high resolution (i.e., with no cross-coupling between the separated modes). A typical way to achieve mode sorting is based on the projective measurement, which can be easily implemented with a hologram \cite{Zhang:10}, but the efficiency is inherently limited to $1/N$ for sorting $N$ modes, which does not scale well with the size of the OAM state space. The low-efficiency problem also exists in OAM sorting based on integrated photonic devices due to the limited coupling area and complicated fabrication \cite{Su:12}. Alternative approaches including interferometric methods \cite{PhysRevLett.88.257901} (similar design are also demonstrated for sorting radial modes very recently in Ref.\cite{PhysRevLett.119.263602}) and multi-plane light conversion technique \cite{Labroille:14} can theoretically achieve unity efficiency, while requiring cascading multiple optical elements and the quantity of these elements also increases with the number of sorted modes, making it difficult and complicated to scale up. 

An efficient yet much simpler scheme is based on optical coordinate transformation, which can also achieve unity efficiency and only requires two phase masks. A log-polar transformation is well known for OAM mode sorting \cite{PhysRevLett.105.153601}, in which, log-polar coordinates in the input plane are conformally mapped to Cartesian coordinates in the output plane \cite{doi:10.1080/09500348714551121}. The corresponding optical setup consists of an input phase mask which implements the transformation, and a phase-correction output mask which collimates the beam. After propagating through this system, an input OAM mode with a spiral wavefront $\exp(i\ell \theta)$ is transformed to a tilted plane wavefront $\exp(i\ell x/ \beta)$, where $x$ is the Cartesian coordinate and $\beta$ is a scaling parameter. OAM modes with different $\ell$ are hence transformed to plane wavefronts with different tilt angles which can be focused to distinct positions in the focal plane of a lens, allowing them to be detected and processed as separate channels, as illustrated in Fig.~\ref{fig1}(a). Experimental implementation of this scheme is continuously improved from spatial light modulators (SLMs) \cite{PhysRevLett.105.153601} to custom refractive elements \cite{Lavery:12} and recently to more compact mode sorters with advanced electron beam-lithography \cite{Ruffato:17} and three-dimensional (3D) laser printing techniques \cite{Lightman:17}.

However, the simplicity of this OAM sorting principle based on the log-polar transformation comes at a cost of an inherent limitation in adequately separating adjacent OAM states, which is due to a significant overlap between their distribution in the output plane. The origin of this overlap is in the mathematical transformation itself. For an input OAM mode with topological charge $\ell$, the $2\pi \ell$ excursion of the phase around a complete circle in the input plane is transformed to an equal phase excursion along a line segment in the output plane. After focusing, the cross-section of the field amplitude profile in the Fourier plane resembles a $sinc$ function that is shifted from the center by a wave number proportional to $\ell$. With Fourier analysis, the position of the $sinc$ function equals ${k_\ell } = \ell /b$ while its width is $2/b$ (defined as the distance between the two first zeros either side of the central peak). The peaks of two adjacent OAM modes with topological charges $\ell ,\ell  + 1,$ can therefore be separated by ${k_{\ell  + 1}} - {k_\ell } = 1/b,$ which is only half the width of the $sinc$ functions, resulting in significant power overlap between them and thus in limited discrimination resolution giving rise to cross-talk between the separated channels, as shown in Fig.~\ref{fig1}(a). To alleviate this problem, a beam-copying technique can be employed, but at the cost of a more complex optical setup and additional numerical effort for optimization \cite{OSullivan:12,Nat.Common.4.2781,doi:10.1063/1.4974824}.

In this Letter, we take a fundamentally different approach to overcome the limitations of the log-polar OAM mode sorting scheme. We propose that spiral transformations mapping spirals (instead of closed circles) to parallel lines could be exploited for OAM mode sorting with higher modal resolution. As illustrated in Fig.~\ref{fig1}, compared to the log-polar method, the spiral transformation scheme provides naturally an $n$-fold increase of the phase excursion along the output wavefront, where the effective number $n$ of the spiral turns being imaged is limited only by the beamwidth of the input OAM mode. The extended output wavefront is then mapped in the Fourier plane to a $sinc$ function which is also shifted by ${k_\ell } = \ell /\beta$ but with a narrower width that is inversely proportional to the number of turns, $2/(n\beta )$ which translates to a significantly reduced overlap between adjacent OAM modes. With this idea, we show in the theoretical analysis that there could exist different types of spiral transformations. The simplest analytical transformation among these that conformally maps logarithmic spirals to parallel lines is specifically demonstrated, which is also found to generalize the standard log-polar transformation.  These theoretical predictions are further verified through numerical simulation and experiment.

\begin{figure}
{\centerline{\includegraphics[width=8.5cm]{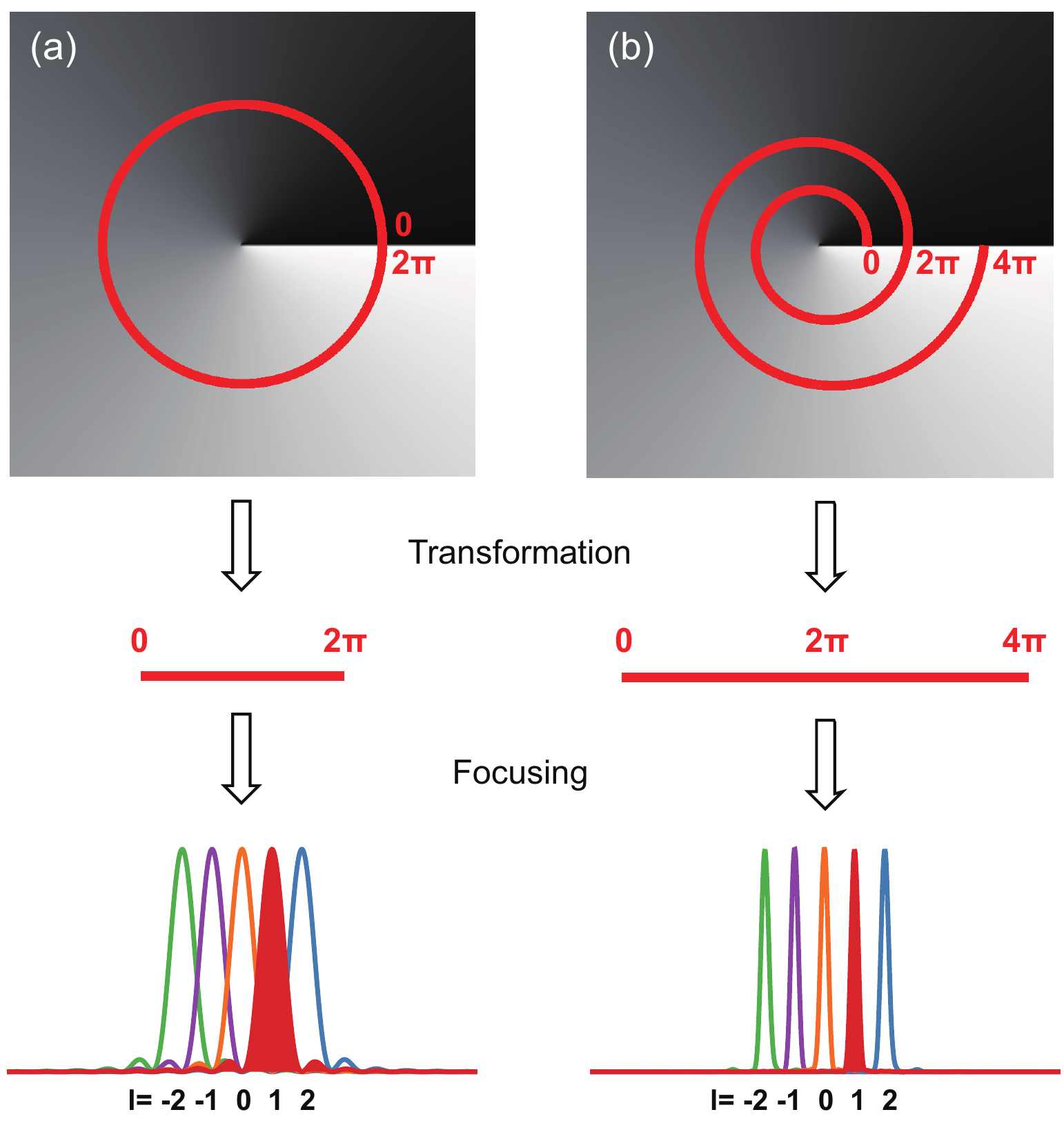}}}
\caption{\label{fig1}Principle of the (a) log-polar transformation scheme, and (b) the introduced spiral transformation scheme for OAM mode sorting.}
\end{figure}

In order to obtain the spiral coordinate transformation, we consider paraxial propagation of a light wave between an input plane $(x,y)$ and an output plane $(u,v)$ that is parallel to the input plane and located at the distance $d$. The mapping of point $(x,y)$ to point $(u,v)$ is given in the context of ray optics by the equations
\begin{equation}
{Q_x} = k\frac{{u - x}}{d},\quad {Q_y} = k\frac{{v - y}}{d},
\label{eq:1}
\end{equation}
where $Q(x,y)$ is the phase distribution in the input plane (imposed in practice by a phase mask), ${Q_x}$ and ${Q_y}$ being its partial derivatives with respect to the corresponding variables, and $k$ is the free-space wavenumber. We now introduce the new coordinates $(s,\theta)$ in the $(x,y)$ plane according to the equations
\begin{equation}
x = r\left( {s,\theta } \right)\cos \theta ,\quad y = r\left( {s,\theta } \right)\sin \theta,
\label{eq:2}
\end{equation}
where $(r,\theta)$ are the polar coordinates of the point $(x,y)$. The new coordinates $(s,\theta)$ can be considered as spiral-polar coordinates, where the parameter $s$ indicates the particular spiral that the point $(x,y)$ belongs to and the polar angle $\theta$ determines the position of the point on that spiral. Playing the role of a position parameter along a spiral of infinite length, the angle $\theta$ varies without restriction in $\left[ {0,\infty } \right)$, which is already the first difference with the standard log-polar transformation where the angle is limited to a $\left[ {0,2\pi } \right)$ variation. The shape of the spiral is determined by the function $r\left( {s,\theta } \right)$. For example, an Archimedean spiral can be expressed as $r\left( {s,\theta } \right) = s + a\theta$ and a logarithmic spiral as $r\left( {s,\theta } \right) = s\cdot\exp (a\theta )$, $a>0$. 

Now consider the mapping of an input spiral $r(s,\theta)$ to an output straight line, which can be expressed as
\begin{equation}
u = u\left( {s,\theta } \right),\quad v = v\left( s \right).
\label{eq:3}
\end{equation}
These equations imply that the spiral labeled by the variable $s$ can be mapped to a horizontal line in the output plane (parallel to the $u$ axis) with the position $u$ along that line generally depending on both $s$ and $\theta$. Spirals with different parameter $s$ are therefore mapped to parallel horizontal lines displaced in the vertical direction (parallel to the $v$ axis) according  to the function $v(s)$. 

If the input phase distribution $Q(x,y)$ is to be a twice continuously differentiable function, we then have ${Q_{xy}} = {Q_{yx}}$. From Eq.~(\ref{eq:1}), we have ${u_y} = {v_x}$, and due to Eq.~(\ref{eq:3})
\begin{equation}
{u_s}{s_y} + {u_\theta }{\theta _y} = v'\left( s \right){s_x},
\label{eq:4}
\end{equation}
which can be further rewritten in polar coordinates as
\begin{equation}
\left[ {{u_s}{r_\theta } - {u_\theta }{r_s} + v'\left( s \right)r} \right] \cos \theta + \left[ {v'\left( s \right){r_\theta } - {u_s}r} \right] \sin \theta  = 0.
\label{eq:5}
\end{equation}

\noindent This is the general condition imposed on the spiral transformation, which could lead to a number of solutions of $r(s,\theta)$ corresponding to different types of spirals. In the following, we are going to impose more restrictions for simplification and obtain one of the simplest cases for demonstration. Since Eq.~(\ref{eq:5}) should be satisfied for all $\theta$ along the spiral, the most obvious way to obtain such an identity is to require that both factors of the trigonometric functions equal zero, which can lead to 
\begin{equation}
{u_s} = {u_\theta }\frac{{{r_s}{r_\theta }}}{{{r^2} + r_\theta ^2}},\quad v'\left( s \right) = {u_\theta }\frac{{{r_s}r}}{{{r^2} + r_\theta ^2}}.
\label{eq:6}
\end{equation}
\noindent For a given spiral shape $r(s,\theta)$, Eqs.~(\ref{eq:6}) relate the functions $v'\left( s \right)$, ${u_s}$ and ${u_\theta }$ which have to be solved in order to yield the image $(u,v)$ of the spiral in the output plane. However, since there are only two equations, one of the three unknown functions has to be determined by a separate requirement. For the OAM mode sorting purpose, it is desired to map the constant phase gradient along the azimuth of an input OAM mode, namely $\exp(i \ell \theta)$ to a constant phase gradient along the $u$ direction of the output mode, namely $\exp(i \ell u/\beta)$. Hence a linear relationship between $\theta$ and $u$ is assumed, or  ${u_\theta } = \beta$, where constant $\beta$ is the scaling parameter mentioned before. Under this assumption, the fraction about $r$ in the right-hand side of the second sub equation of Eqs.~(\ref{eq:6}) should be independent of $\theta$, which obviously leads to a simple solution of a logarithmic spiral $r\left( {s,\theta } \right) = s\cdot\exp (a\theta )$. Equations~(\ref{eq:6}) can then be integrated to yield
\begin{equation}
u\left( {s,\theta } \right) = \frac{{a\beta }}{{1 + {a^2}}}\ln \left( {\frac{s}{{{r_0}}}} \right) + \beta \theta ,\quad v\left( s \right) = \frac{\beta }{{1 + {a^2}}}\ln \left( {\frac{s}{{{r_0}}}} \right),
\label{eq:7}
\end{equation}
where we have arbitrarily chosen that the point $\theta  = 0$ of the spiral with $s = {r_0}$ (an arbitrary positive constant) is mapped to $(0,0)$. The transformation can be rewritten in the form of polar coordinates $(r,\theta)$ in the input plane 
\begin{equation}
\begin{split}
u\left( {r,\theta } \right) = \frac{\beta }{{1 + {a^2}}}\left[ {a\ln \left( {\frac{r}{{{r_0}}}} \right) + \theta } \right] \\
v\left( {r,\theta } \right) = \frac{\beta }{{1 + {a^2}}}\left[ {\ln \left( {\frac{r}{{{r_0}}}} \right) - a\theta } \right]
\end{split}~.
\label{eq:8}
\end{equation}

\noindent It should be noted that the polar angle $\theta$ here is not limited to $\left[ {0,2\pi } \right)$ but $\left[ {0,\infty } \right)$ as mentioned before, which plays the role of the position parameter of a spiral as shown in Supplemental Material (SM) \cite{supplement}.
 
Equation~(\ref{eq:8}) describes the simplest spiral transformation to be specifically demonstrated in this Letter. It is also interesting to find that this spiral transformation actually generalizes the standard log-polar mapping from a mathematical viewpoint. Indeed, by introducing the complex variables $W = v + iu$ and $Z = x + iy$, Eq.~(\ref{eq:8}) can be written compactly as the conformal mapping between the $Z$ (input) and $W$ (output) complex planes 
\begin{equation}
W = \beta {e^{i\phi }}\ln \left( {\frac{Z}{{{r_0}}}} \right),
\label{eq:9}
\end{equation} 

\noindent where $\phi  = ta{n^{ - 1}}\left( a \right)$.  By setting $a=0$, we have $\phi=0$ and Eq.~(\ref{eq:9}) reduces to the standard log-polar transformation. Figure~\ref{fig2} depicts this new transformation as a conformal mapping between the input and output coordinates. A continuous bundle of spirals with $s$ taking values in an interval $[{s_{\min }},{s_{\max }}]$ is mapped to a horizontal strip that is slightly deformed (further analysis in SM \cite{supplement}). For this continuous bundle of spirals, the maximum allowed value of the ratio ${{{s_{\max }}} \mathord{\left/{\vphantom {{{s_{\max }}} {{s_{\min }}}}} \right.\kern-\nulldelimiterspace} {{s_{\min }}}}$ is $\exp \left( {2\pi a} \right)$, which is the extreme case when the last spiral of the bundle (spiral with $s = {s_{\max }}$) becomes identical with the second turn of the first spiral of the bundle (spiral with $s = {s_{\min }}$). This condition ensures that the strip formed by the bundle of spirals covers the entire input plane without leaving gaps between successive turns and hence will be employed later, while a smaller ratio is assumed in Fig.~\ref{fig2} for better illustration.

\begin{figure}
{\centerline{\includegraphics[width=8.5cm]{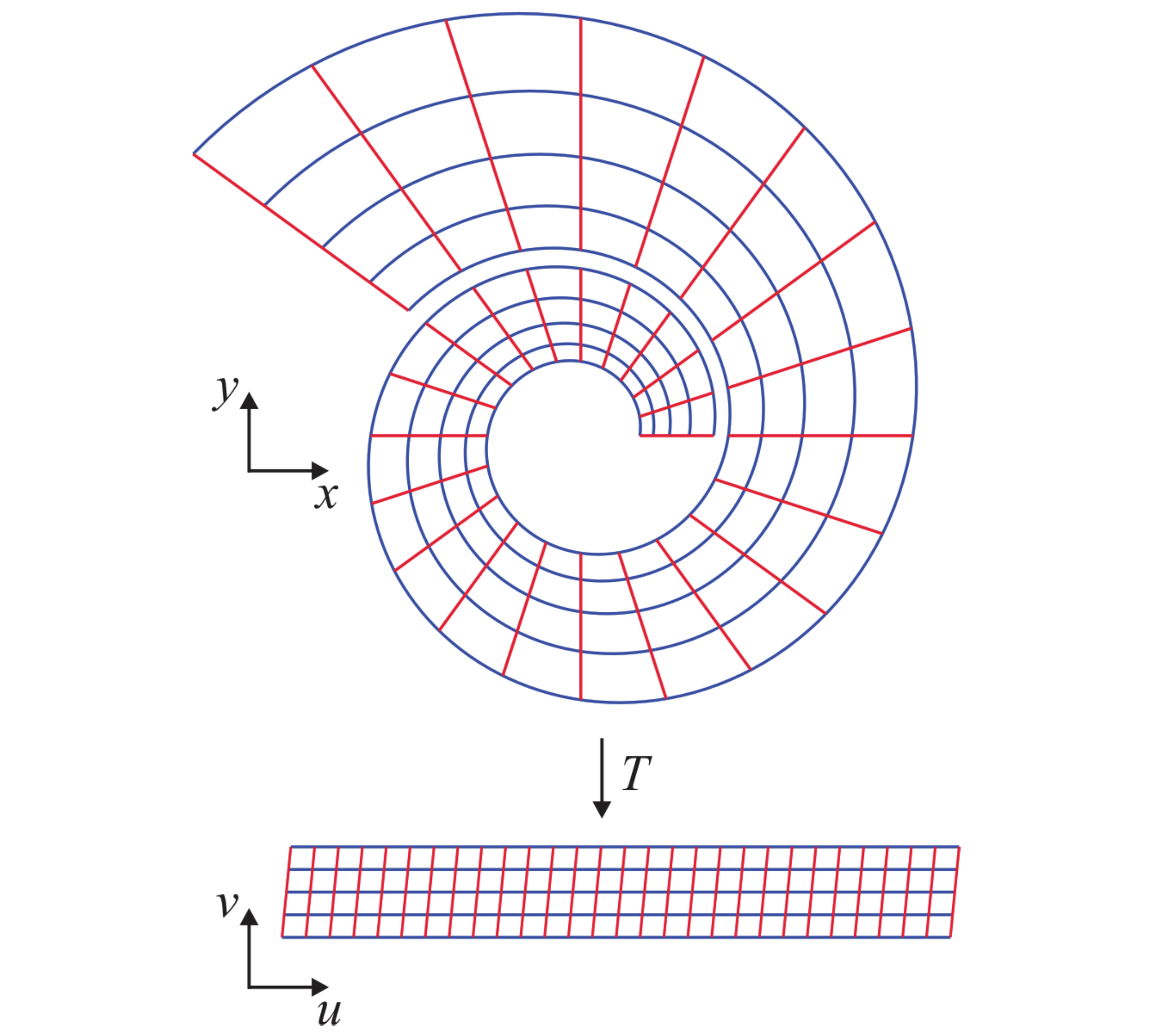}}}
\caption{\label{fig2}The spiral transformation (indicated with $T$) as a conformal mapping between the planes $(x,y)$ and $(u,v)$. Blue lines indicate the spirals (curves of constant $s$) and red lines indicate the azimuthal positions (lines of constant $\theta$).}
\end{figure}

Having derived the form of a spiral transformation, the input phase $Q(x,y)$ implementing this transformation can be obtained based on Eqs.~(\ref{eq:1}) and (\ref{eq:8}) as 
\begin{equation}
Q = \frac{{k\beta }}{{d\left( {{a^2} + 1} \right)}}\left[ \begin{array}{l}
\left( {ax + y} \right)\ln\left( {\frac{r}{{{r_0}}}} \right) + \\
\left( {x - ay} \right)\theta  - \left( {ax + y} \right)
\end{array} \right] - \frac{{k{r^2}}}{{2d}}.
\label{eq:10}
\end{equation} 

\noindent This is the input phase required to map spirals of the input plane to parallel lines in the output plane. The phase of this output mode will contain two terms. The first term is the linear phase gradient transformed from the angular phase gradient of the input OAM mode along the mapped spiral paths, which enables OAM mode sorting in the Fourier plane. The second term is the phase acquired by wave propagation between the planes $(x,y)$ and $(u,v)$, which should be compensated with a phase-correction mask. If $P(u,v)$ is the phase distribution of this mask, its local gradient at a point $(u,v)$ should cancel the slope of the ray arriving at this point, which reads
\begin{equation}
{P_u} = k\frac{{x - u}}{d},\quad {P_v} = k\frac{{y - v}}{d}.
\label{eq:11}
\end{equation}  

\noindent After substituting $(x,y)$ as functions of (u,v) based on Eq.~(\ref{eq:8}), Eq.~(\ref{eq:11}) can be integrated to obtain
\begin{equation}
\begin{split}
P = & \frac{{k{r_0}}}{d}\frac{\beta }{{1 + {a^2}}}\exp\left( {\frac{{au + v}}{\beta }} \right) \times \left[ {\sin\left( {\frac{{u - av}}{\beta }} \right)} \right. \\
& \left. { + a\cos\left( {\frac{{u - av}}{\beta }} \right)} \right] - \frac{{k\left( {{u^2} + {v^2}} \right)}}{{2d}}
\end{split}~.
\label{eq:12}
\end{equation}

To verify our theoretical predictions, numerical simulation and experiment are implemented to confirm the performance of the spiral transformation in sorting OAM modes compared with the log-polar transformation. The numerical simulation is performed based on the angular spectrum diffraction theory with Laguerre-Gaussian beams as the input OAM modes. The parameters in the transformations are chosen to ensure diffraction within the paraxial regime. Typical values used are $d = 134{\rm{ }}$ mm, $2\pi \beta {\rm{ = }}2{\rm{ }}$ mm and $2\pi a = ln(1.6)$, at a telecommunication wavelength $\lambda  = 1550{\rm{ }}$ nm.

\begin{figure}
{\centerline{\includegraphics[width=8.5cm]{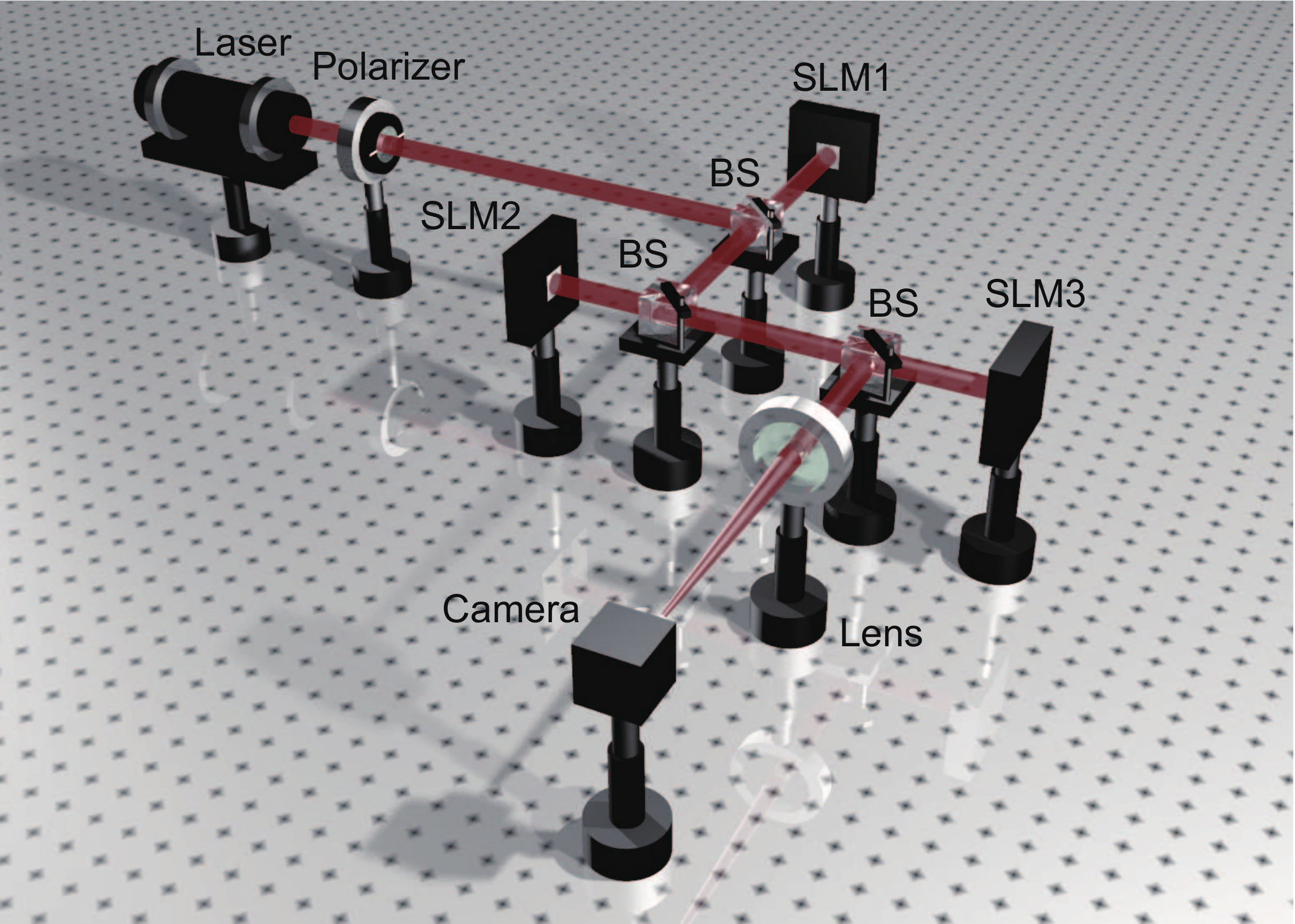}}}
\caption{\label{fig3}Schematic of the optical setup. BS: beam splitter.}
\end{figure}

The corresponding experimental setup is shown in Fig.~\ref{fig3}. A laser operating at a wavelength of $1550$ nm emits a collimated Gaussian beam with a spot size of around $2$ mm in diameter, which is horizontally polarized by a polarizer to match the polarization response of the SLMs. The first SLM (SLM1) is programmed to produce a vortex beam with the desired beam width and topological charge. The generated OAM state is subsequently reflected on SLM2 which reproduces the phase distribution $Q(x,y)$ given by Eq.~(\ref{eq:10}) and shown in SM \cite{supplement}. This phase mask is termed as the unwrapper and is responsible for implementing the spiral or log-polar transformation. After propagation for a certain distance, the beam reflected by SLM2 illuminates SLM3 which implements the phase distribution $P(x,y)$ given by Eq.~(\ref{eq:12}) and works as the phase corrector. After it is reflected by SLM3, the beam has a strip-shaped transverse profile and a linear phase gradient transformed from the input vortex beam. This beam is finally Fourier transformed at the focal plane of a lens, where it is captured by an infrared camera. The parameters used in the experiment are consistent with the simulation.

\begin{figure}
{\centerline{\includegraphics[width=8.5cm]{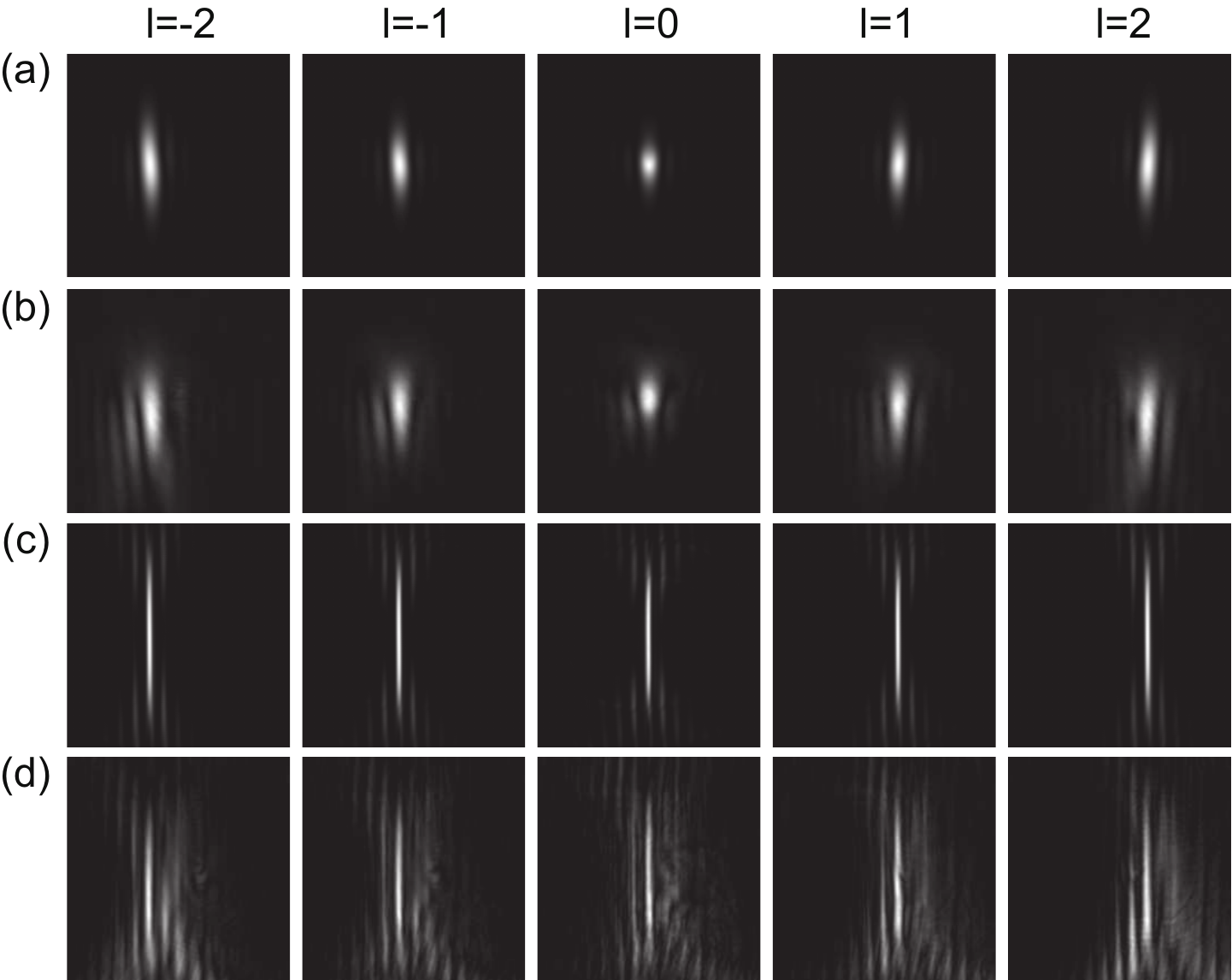}}}
\caption{\label{fig4}Numerical (a, c) and experimental (b, d) intensity distributions of the eventually separated vortex beams with topological charges $- 2 \le l \le 2$ based on the log-polar (a, b) and the spiral (c, d) transformation scheme. All images refer to the focal plane of the lens (Fourier plane) and have the same spatial scale.}
\end{figure}

\begin{figure}
{\centerline{\includegraphics[width=8.5cm]{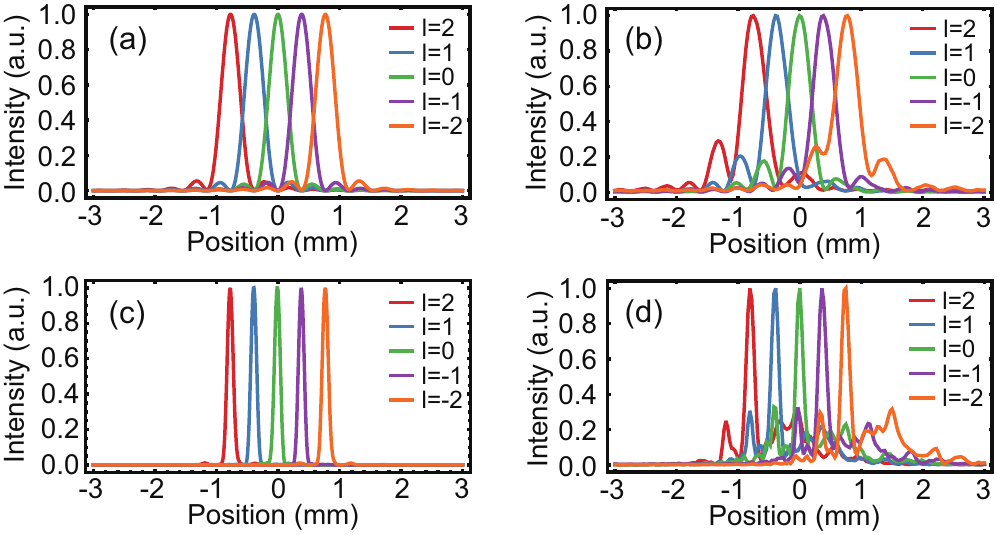}}}
\caption{\label{fig5}Intensity profiles along a horizontal cut of images in Fig.~\ref{fig4} for different separated OAM modes. (a) Numerical and (b) experimental results based on the log-polar transformation scheme; (c) numerical and (d) experimental results based on the spiral transformation scheme.}
\end{figure}

Figure~\ref{fig4} shows the numerical and experimental results of the eventually separated OAM modes $\left( { - 2 \le \ell  \le 2} \right)$ based on both transformation schemes (the transformed beams shown in SM \cite{supplement}), and Fig.~\ref{fig5} depicts the intensity profiles along a horizontal cut of these separated OAM modes. As expected for both schemes, OAM modes with different topological charges are horizontally displaced from the center by a shift proportional to their topological charge. The shift is the same in both schemes for modes with the same topological charge, because both schemes are designed to share the same scaling parameter $\beta$ ($2\pi \beta {\rm{ = }}2{\rm{ }}$ mm). For a focal distance $f = 0.5{\rm{ }}$ m, the theoretical spacing between adjacent OAM modes is ${{\lambda f} \mathord{\left/{\vphantom {{\lambda f} {\left( {2\pi \beta } \right)}}} \right.\kern-\nulldelimiterspace} {\left( {2\pi \beta } \right)}}{\rm{ = }}387.5{\rm{ }}$ $\mu$m  which is very close to the experimentally measured value in Fig.~\ref{fig5}.
 
What is, however, most important about the results is the significant improvement in the separation of the OAM modes by the spiral transformation compared to the log-polar one. This is obvious either by comparing the numerical results of Fig.~\ref{fig5}(a) and ~\ref{fig5}(c) or by comparing the experimental results of Fig.~\ref{fig5}(b) and ~\ref{fig5}(d), where the $sinc^2$ profiles of the separated OAM modes in the spiral scheme are clearly narrower and less overlapping than the corresponding profiles of the log-polar scheme. The improvement can be quantified by using the concept of optical finesse which is defined as the ratio of the spacing between adjacent OAM states over their average full width half maximum (FWHM) \cite{doi:10.1063/1.4974824}. In our case, this indicator is approximately tripled, from 1.13 for the log-polar scheme to 3.48 for the spiral scheme in simulation and from 0.95 to 2.60 in experiment, and the slight difference between numerical and experimental values results from beam expansion in experiment. It is also noted that experimental results seem to be affected from some stray light in the background, which is attributed to small errors in the 3D alignment of the SLMs, such as transverse displacement, rotation and distance between them, leading to imperfect phase correction in the output plane. Such problems can be overcome if the entire sorting scheme is arranged in a single device that comprises both the unwrapper and the phase corrector \cite{Ruffato:17,Lightman:17}.

In summary, we have proposed a novel spiral transformation scheme for high-resolution OAM mode sorting. Theorectically, there exist different types of spiral transformations, one of which specifically demonstrated in this Letter maps logarithmic spirals to parallel lines and it was interesting to find that this particular spiral transformation generalizes the well-known log-polar transformation.  The transformation was derived analytically starting from first principles and the analysis concluded with the computation of the phase masks required to implement this conformal mapping. The theoretical predictions were further verified through numerical simulation and experiment, with an expected obvious reduction in the overlap between adjacent sorted OAM modes, compared to the log-polar transformation scheme.

The new optical transformation scheme demonstrated in this work widens our views of the mathematical tools and concepts available toward efficient OAM mode sorting. Further possibilities and more sophisticated transformation schemes might emerge if similar concepts are sought beyond the regimes of ray optics and paraxial propagation, as for example in the context of wave optics, optical transformation media and metamaterials.

$\\$
\indent This work is supported by the National Basic Research Program of China (973 Program) (2014CB340000), National Natural Science Foundation of China (NSFC) (61490715, U1701661, 11774437, 61323001, 11690031), Science and Technology Program of Guangzhou (201707020017, 2018-1002-SF-0094), EU
H2020 project ROAM, and Fundamental Research Funds for the Central Universities of China (SYSU: 17lgzd06).  EU H2020 project ROAM.  The authors thank Guoxuan Zhu (SYSU) for helpful discussion with the experiment. 


%

\clearpage

\textbf {Supplemental Material}

\section{Further analysis on the spiral transformation}

Noted that in Eq. (8) of the main text, due to the dependence of $u$ on $s$, lines of constant azimuth $\theta$ in the input plane are mapped to lines in the output plane that are not vertical to the $u$ axis but tilted. Another way to express this is that spirals with different $s$ values are mapped to lines that are relatively displaced along the $u$ axis. As a result, a continuous bundle of spirals with $s$ taking values in an interval $[{s_{\min }},{s_{\max }}]$ is mapped to a horizontal strip that is not exactly rectangular but is deformed by a horizontal shear stress, as shown in Fig. 2 of the main text. The effect of this stress on the Fourier transform of the output mode can be deduced from Eq. (8) in the main text which are combined to give $\beta \theta  = u - av$. The vortex phase $\exp(i\ell \theta )$ is thus mapped to a plane wavefront with phase $\exp(i\ell (u - av)/\beta )$ which reveals the existence of an additional wave vector $a\ell /\beta$ along the direction $v$. This has the simple consequence that the focused mode will be centered at the off-axis position $\left( {\ell /\beta , - a\ell /\beta } \right)$ in the Fourier plane. This however does not affect the sorting process since the focused mode is already extended in the $v$ direction due to the narrow width of the transformed strip-shaped mode in the same direction. Furthermore, this vertical displacement in the Fourier space will be small compared to the horizontal displacement because a small growth rate $(a<<1)$ is preferred in practice so that the logarithmic spirals complete several turns within the finite input aperture.

\section{Supplemental figures}

\subsection{Description of the spiral with polar coordinates}

Noted that the polar coordinates $(r,\theta)$ used to describe the spiral and also the phase distributions performing the corresponding spiral transformation are not standard polar coordinates, in which the polar angle $\theta$ is not limited to $\left[ {0,2\pi } \right)$. Instead, it can be written as $\theta  = {\theta _0} + 2m\pi$, where $\theta _0$ is the standard polar angle and $m= 0,1,...$ corresponding to the first, second, etc. turns of the spiral, as illustrated in Fig. S1. The relationship of this polar coordinates $(r,\theta)$ and the Cartesian coordinates $(x,y)$ can be expressed as

\begin{equation}
r = {({x^2} + {y^2})^{1/2}},\   \theta  = {\theta _0} + 2m\pi,  \tag{SM1}
\label{eq:9}
\end{equation} 

\noindent where
\begin{equation}
\begin{split}
{\theta _0} = ta{n^{ - 1}}\left( {\frac{y}{x}} \right) \in \left[ {0,2\pi } \right) \\
m = \left\lfloor {\frac{1}{{2\pi a}}ln\left( {\frac{r}{{{r_0}}}{e^{ - a{\theta _0}}}} \right)} \right\rfloor 
\end{split}~, \tag{MS2}
\label{eq:8}
\end{equation}

\noindent where $\left\lfloor {\ } \right\rfloor$ stands for the integer part.

\subsection{Phase masks programmed on the SLMs}

Figure S2 shows the mod-$2\pi$ phase distribution of the unwrapper and the phase corrector for both the standard log-polar transformation and the spiral transformation proposed in this paper, which are depicted according to the derived analytical expressions of Eqs. (10) and (12) in the main text. These images are then loaded to the SLM2 and SLM3 in experiment to implement the corresponding phase modulation.

\subsection{Transformed beams after the log-polar and spiral transformation}

Figure S3 summarizes the numerical and experimental results for vortex beams with different topological charges being transformed by the log-polar [Figs. S3(a) and S3(b)] and the spiral [Figs. S3(c) and S3(d)] transformation principle. Note that, in the standard log-polar scheme, the input OAM modes are transformed to rectangular stripes of limited length $2\pi \beta {\rm{ = }}2{\rm{ }}$ mm, according to our selection for the scaling parameter $\beta$. On the contrary, in the spiral transformation scheme, the OAM modes are mapped to significantly longer lines which reflects the number of the spiral turns (about three turns on average in this case) being mapped within the transverse extent of the power distribution of the input vortex mode. The agreement of the experimental results with the numerical simulations, as far as the shape and the length of the transformed images, is very satisfactory. These tripled elongated transformed beams based on the spiral transformation compared to the log-polar transformation result in narrower intensity profiles after focusing as presented in Fig. 4 and Fig. 5 of the main text, which is quantitatively characterized by the optical finesse with approximately three times improvement.

\begin{figure*}
{\centerline{\includegraphics[width=8.5cm]{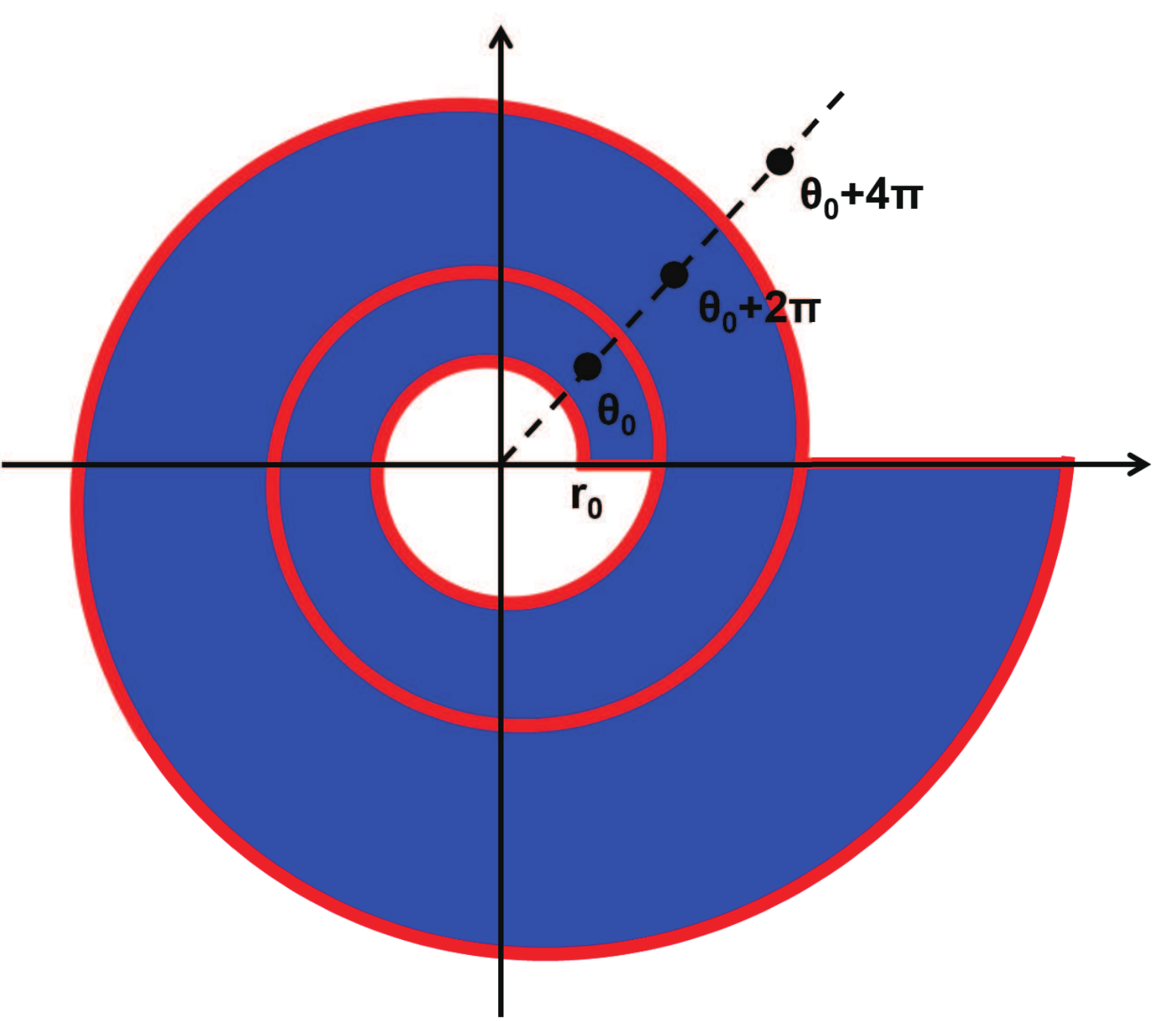}} }
\caption*{\label{fig1} Fig. S1. Schematic of a spiral described by the polar coordinates.}
\end{figure*}

\begin{figure*}
{\centerline{\includegraphics[width=8.5cm]{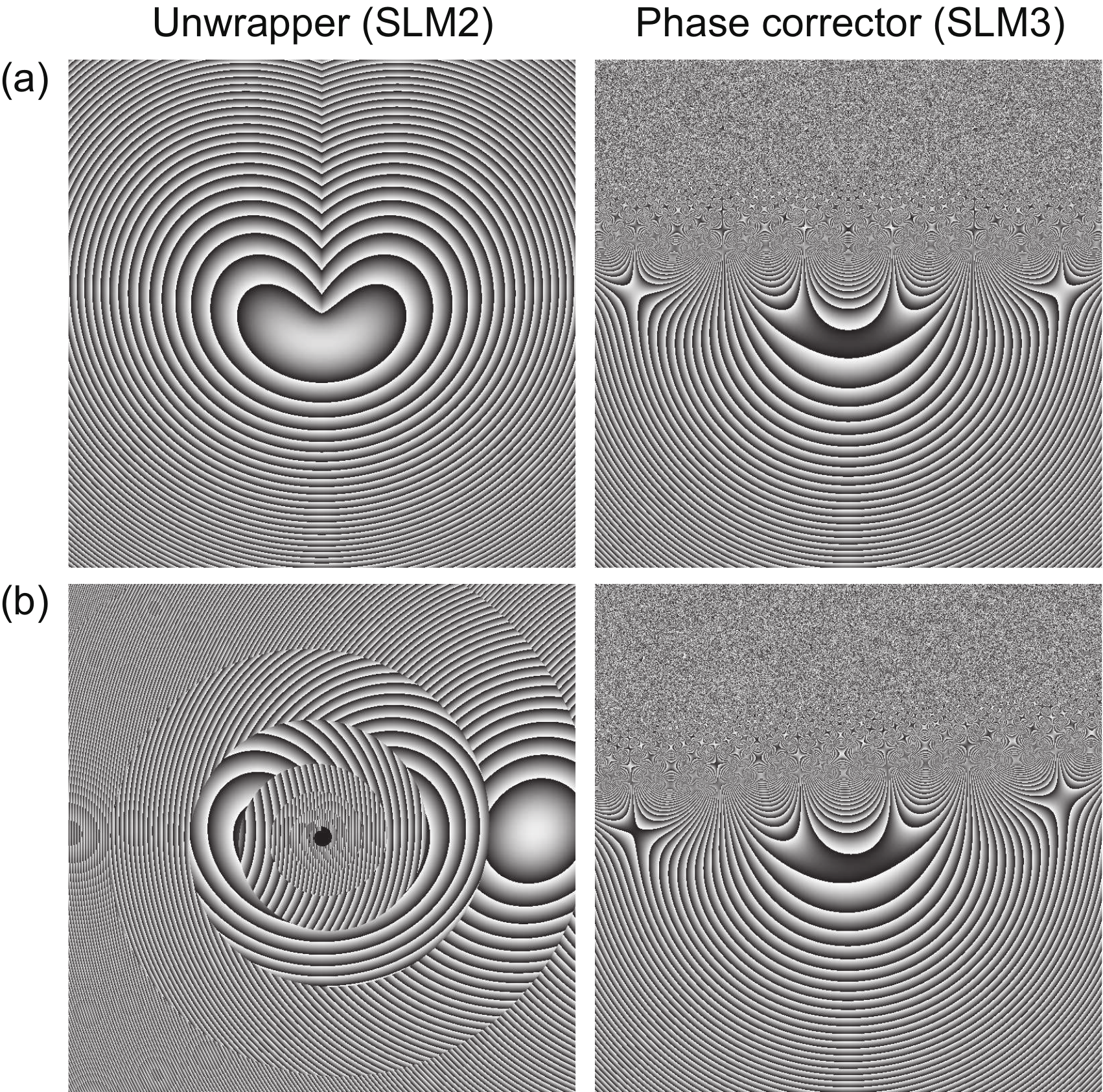}}}
\caption*{\label{fig2}Fig. S2. Phase masks programmed on the SLM2 and SLM3 corresponding to the unwrapper (left column) and the phase corrector (right column) for the log-polar (row $a$) and the spiral (row $b$) transformation scheme. All images use the same scale.}
\end{figure*}

\begin{figure*}[b]
{\centerline{\includegraphics[width=16cm]{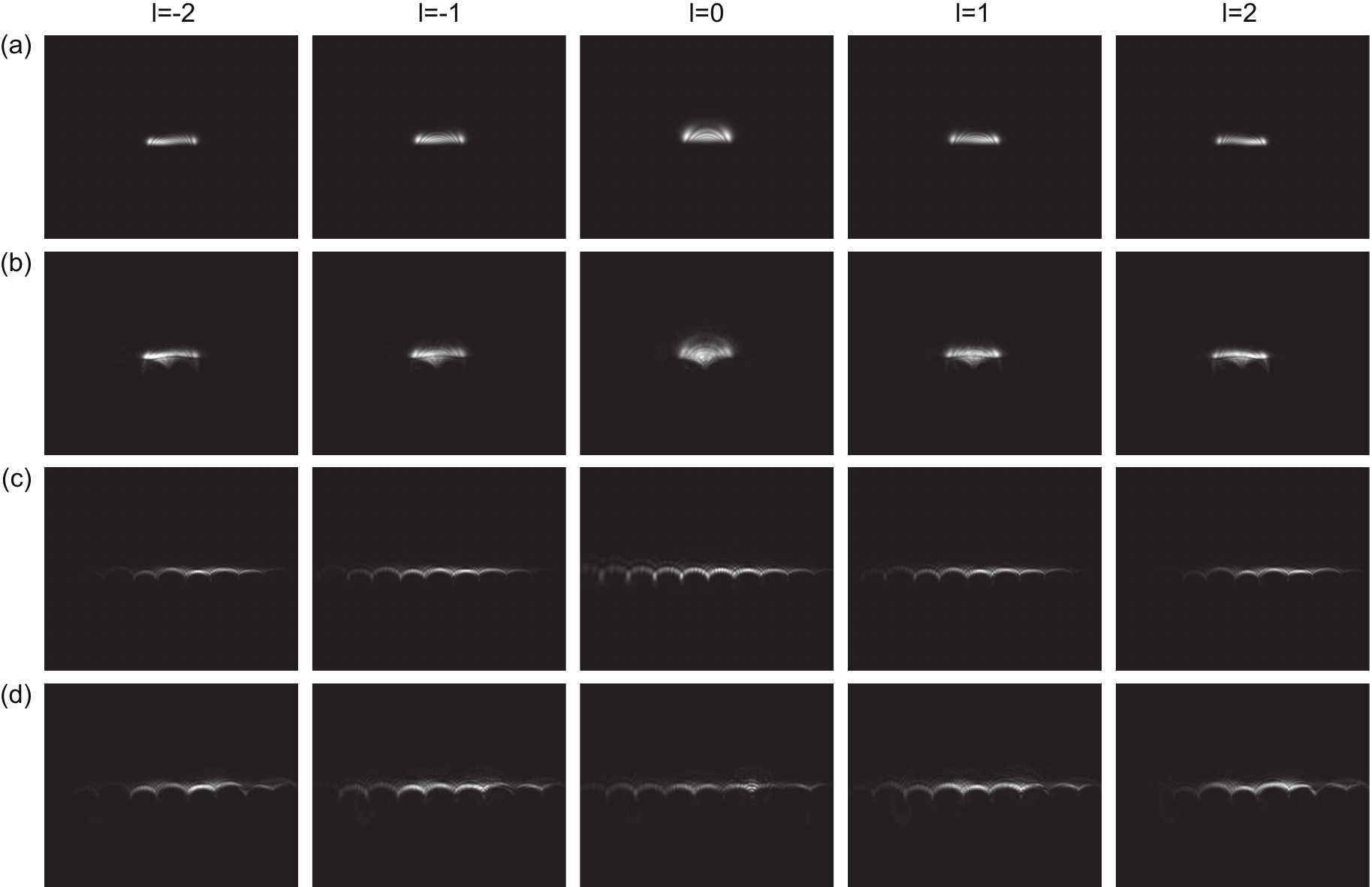}}}
\caption*{\label{fig3}Fig. S3. Numerical (a, c) and experimental (b, d) intensity distributions of vortex beams with topological charges $- 2 \le l \le 2$ transformed by the log-polar (a, b) and the spiral (c, d) transformation scheme. All images refer to the plane just after (or just before) the phase corrector (SLM3) and have the same spatial scale.}
\end{figure*}

\end{document}